%% file: MAIN.tex
\documentclass[a4paper,twoside]{article}
\usepackage{epsfig}
\usepackage{subcaption}
\usepackage{calc}
\usepackage{amssymb}
\usepackage{amstext}
\usepackage{amsmath}
\usepackage{amsthm}
\usepackage{multicol}
\usepackage{apalike}
\usepackage{algorithm2e}
\usepackage[bottom]{footmisc}

\usepackage{bm}
\usepackage{enumitem}
\usepackage{multicol}
\usepackage{graphicx}
\usepackage{siunitx}
\usepackage{rotating}
\usepackage{tablefootnote}

\usepackage{mathptmx}
\usepackage{newtxmath}
\usepackage[colorlinks=true, linkcolor=black, urlcolor=black, citecolor=black]{hyperref}
\graphicspath{{./figures/}}

\usepackage{SCITEPRESS} 

\begin{document}
{\LARGE Copyright Notice}
\newline
\fboxrule=0.4pt \fboxsep=3pt
\fbox{\begin{minipage}{1.1\linewidth}
		\textcopyright\,\,2024\, International Conference on Informatics in Control, Automation and Robotics (ICINCO)\\ \\
		This is the author's version of an article that has been accepted to ICINCO 2024 (SciTePress proceedings). Changes will be made to this version by the publisher prior to publication.\\ \\
        The proceedings version will be distributed as open access under CC BY-NC-ND 4.0 license. No further permission to reuse this article is required. For other type of use, contact SciTePress, Lda.
\end{minipage}}

\title{Domain-decoupled Physics-informed Neural Networks with Closed-form Gradients for Fast Model Learning of Dynamical Systems}

\author{\authorname{Henrik Krauss\sup{1,2}\orcidAuthor{0000-0002-9787-5465}, Tim-Lukas Habich\sup{2}\orcidAuthor{0000-0003-4167-8443}, Max Bartholdt\sup{2}\orcidAuthor{0000-0002-8422-9368},\\Thomas Seel\sup{2}\orcidAuthor{0000-0002-6920-1690} and Moritz Schappler\sup{2}\orcidAuthor{0000-0001-7952-7363}}
\affiliation{
\sup{1}Department of Advanced Interdisciplinary Studies, The University of Tokyo, Tokyo, Japan \\
\sup{2}Institute of Mechatronic Systems, Leibniz University Hannover, 30823 Garbsen, Germany
}
\email{henrik-krauss@g.ecc.u-tokyo.ac.jp, \\ \{tim-lukas.habich, 
max.bartholdt, 
thomas.seel, moritz.schappler\}@imes.uni-hannover.de}
}


\abstract{
Physics-informed neural networks (PINNs) are trained using physical equations and can also incorporate unmodeled effects by learning from data.
PINNs for control (PINCs) of dynamical systems are gaining interest due to their prediction speed compared to classical numerical integration methods for nonlinear state-space models, making them suitable for real-time control applications.
We introduce the domain-decoupled physics-informed neural network (DD-PINN) to address current limitations of PINC in handling large and complex nonlinear dynamical systems.
The time domain is decoupled from the feed-forward neural network to construct an Ansatz function, allowing for calculation of gradients in closed form. This approach significantly reduces training times, especially for large dynamical systems, compared to PINC, which relies on graph-based automatic differentiation.
Additionally, the DD-PINN inherently fulfills the initial condition and supports higher-order excitation inputs, simplifying the training process and enabling improved prediction accuracy.
Validation on three systems -- a nonlinear mass-spring-damper, a five-mass-chain, and a two-link robot~-- demonstrates that the DD-PINN achieves significantly shorter training times. In cases where the PINC's prediction diverges, the DD-PINN's prediction remains stable and accurate due to higher physics loss reduction or use of a higher-order excitation input.
The DD-PINN allows for fast and accurate learning of large dynamical systems previously out of reach for the PINC.
}
\onecolumn \maketitle \normalsize \setcounter{footnote}{0} \vfill

\input{chapters/chapter1}
\input{chapters/chapter2}
\input{chapters/chapter3}
\input{chapters/chapter4}
\input{chapters/chapter5}

\section*{\uppercase{ACKNOWLEDGEMENTS}}
This work was partially funded by the German Research Foundation (DFG, project numbers 405032969 and 433586601) and the Lower Saxonian Ministry of Science and Culture in the program zukunft.niedersachsen.
\bibliographystyle{apalike}
{\small
\bibliography{MAIN}}
\onecolumn
\section*{APPENDIX}
\input{tables/network_parameters}
\input{tables/results_tab1a}
\input{tables/results_tab1b}
\end{document}

%% file: chapters/chapter1.tex
\section{INTRODUCTION}
\label{sec_intro}
Physics-informed neural networks (PINNs) have been introduced by~\cite{raissi_physicsinformed_2019} as a machine-learning framework to solve ordinary (ODEs) and partial differential equations (PDEs). On top of supervised learning of a feed-forward neural network (FNN) on data, PINNs introduce a custom physics loss based on the system-governing ODEs or PDEs including boundary/initial conditions. This seamless integration of physical laws into the neural-network training process leads to \emph{accurate predictions even with limited data} as well as the ability to \emph{extra\-polate on out-of-distribution (unseen) data}, while serving as a \emph{fast surrogate model}.
So far, many variants of PINNs have been successfully applied to fields such as thermodynamics, chemistry, material science \cite{karniadakis_physicsinformed_2021,cuomo_scientific_2022}, and dynamical systems, including robotics \cite{hao_physicsinformed_2023}.
PINNs have been adapted for the control of dynamical systems (PINCs) by~\cite{antonelo_physicsinformed_2024}
where the FNN predicts the system's evolution over a short time interval assuming constant excitation. This concept is compatible with and has been applied to state estimation using an Unscented Kalman Filter (UKF)~\cite{decurt_hybrid_2024}, model predictive control (MPC) of robotic systems~\cite{nghiem_physicsinformed_2023}, such as two-link manipulators~\cite {nicodemus_physicsinformed_2022,yang_physicsinformed_2023} and quadrotors~\cite{sanyal_rampnet_2023}.
Here, the PINC outperforms traditional numerical integrators regarding prediction speed of the system dynamics.

However, the training of PINNs can be challenging and many efforts to improve trainability have been developed~\cite{wang_expert_2023}. These include adaptive activation functions~\cite{Jagtap_adaptive_2020}, various loss-balancing techniques such as GradNorm~\cite{chen_gradnorm_2018}, SoftAdapt~\cite{heydari_softadapt_2019}, learning rate-annealing (LRA)~\cite{wang_understanding_2021}, relative loss balancing with random lookback (ReLoBRaLo)~\cite{bischof_multiobjective_2021}, dynamically normalized PINNs (DN-PINNs)~\cite{deguchi_dynamic_2023},
adaptive collocation point sampling strategies~\cite{wu_comprehensive_2023}, as well as training on non-dimensional PDEs~\cite{kapoor_physics_2024}. Also, for dynamical systems with chaotic behaviors, loss functions might need to be reformulated to respect spatio-temporal causality to ensure training convergence \cite{wang_respecting_2022}.

One main point of research for PINNs is the topic of differentiation of the FNN output with respect to the spatial and temporal \textit{domain} inputs, i.e., the independent variables in the ODE or PDE over which the solution is defined. Differentiation is required to evaluate the ODE/PDE and determine the physics loss but is \textit{very computationally expensive}. Training times of the PINC~\cite{antonelo_physicsinformed_2024} can amount to several days or weeks, \textit{rendering the learning of large or complex dynamical systems unfeasible}.
Here, previously developed architectures can be summarized in the five categories of using 
\begin{enumerate}[label=(\roman*), leftmargin=*, noitemsep]
    \item automatic differentiation (AD), \label{cat1}
    \item numeric differentiation,\label{cat2}
    \item a hybrid of \ref{cat1}--\ref{cat2},\label{cat3}
    \item a variational approach, or\label{cat4}
    \item closed-form gradients.\label{cat5}
\end{enumerate}

\ref{cat1} Classical PINNs as formulated by~\cite{raissi_physicsinformed_2019} or~\cite{berg_unified_2018} use graph-based \emph{\underline{a}utomatic differentiation} (a-PINNs) as implemented in PyTorch or TensorFlow to determine the FNN gradients. However, AD is computationally expensive and a-PINNs can only reach high accuracy for large numbers of collocation points. 
\ref{cat2} PINNs based on \emph{\underline{n}umerical differentiation} (n-PINNs) avoid AD by using numerical differentiation on the collocation points. This makes them more robust to a low number of collocation points but may attain lower accuracy than a-PINNs \cite{cuomo_scientific_2022}.
\ref{cat3} Addressing the disadvantages of both approaches,~\cite{chiu_canpinn_2022} have used a coupled-automatic-numerical (CAN) differentiation method and introduced \emph{CAN-PINNs} -- a hybrid approach between a- and n-PINNs. Their residual loss is formulated at locally adjacent collocation points through a Taylor-series scheme that includes derivatives at the collocation points obtained from automatic differentiation. However, they require the selection of an appropriate numerical scheme and therefore cannot be straightforwardly interchanged with a-PINNs.
\ref{cat4} Another approach are \emph{\underline{v}ariational PINNs} (v-PINNs) introduced in~\cite{kharazmi_variational_2019}. Here, a variational loss function is constructed within the Petrov-Galerkin framework. For that, the PINNs' output is numerically integrated and tested in a test space of Legendre polynomials.
Variants of the v-PINN have been developed that divide the domain into subdomains \cite{kharazmi_hpvpinns_2021,liu_cvpinn_2023}
\ref{cat5} Also, it is possible to compute \emph{analytical gradients} without graph-based AD, such as for single hidden-layer networks~\cite{lagaris_artificial_1998}. Alternatively, the FNN does not predict the solution to the ODE or PDE directly but parameters of basis or Ansatz functions that approximate the solution. Examples using this approach are:
\begin{itemize}[noitemsep]
    \item Taylor PINN \cite{zhang_solution_2023},
    \item Polynomial-interpolation PINN \cite{tang_physicsinformed_2023},
    \item Legendre-improved extreme learning machines \cite{yang_neural_2020}, and
    \item Multiwavelet-based neural operators \cite{gupta_multiwaveletbased_2021}.
\end{itemize}

In the context of PINNs for control, to the best of our knowledge, only AD-based architectures such as the PINC have been applied~\cite{antonelo_physicsinformed_2024}, featuring the assumption of constant excitation in the short prediction horizon.
Three main limitations of this PINC for the learning of large and nonlinear dynamical systems are:
\begin{enumerate}[label=(\Roman*)]
    \item \label{lim1}
    Graph-based automatic differentiation causes \emph{long training times}. The number of AD operations $n_\mathrm{AD}$ per epoch is proportional to the number of system states $m$ and the number of collocation points $n_\mathrm{collo}$, where $n_\mathrm{collo}$ itself may be chosen higher for larger or more complex systems.
    \item \label{lim2}
    Constant excitation (zero-order-hold assumption) for longer trained time intervals results in \emph{poor accuracy}. However, longer time intervals are advantageous for fast simulation since one prediction with the surrogate model should simulate a large time step with high accuracy.
    \item \label{lim3}
    Initial-condition (IC) loss and physics loss can \emph{require a weighting scheme} such as GradNorm, Soft\-Adapt, LRA or ReLoBRaLo. These weighting schemes need to be chosen empirically, are sensitive to their hyper-parameter settings, and can further increase training time.
\end{enumerate}
For the prediction horizons used in control scenarios, approximating the solution through differentiable Ansatz functions to calculate gradients in closed-form has the potential to drastically reduce training times while maintaining high accuracy.
\begin{figure*}[ht]
	\centering
	\resizebox{1\linewidth}{!}{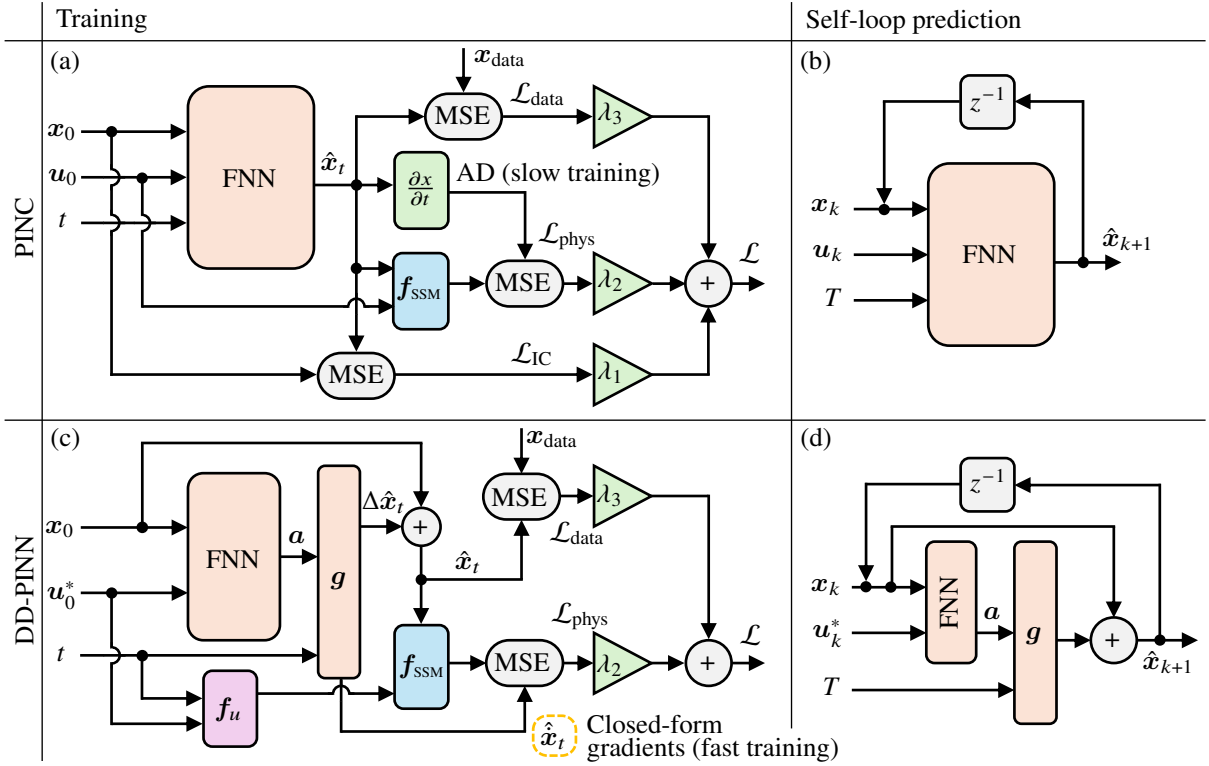}
	\caption{Configurations for (a) PINC training and (b) PINC self-loop prediction process as well as (c) DD-PINN training and (d) DD-PINN self-loop prediction. Time domain is decoupled from the FNN for the DD-PINN, enabling calculation of gradients in closed form.}
    \label{fig_DDPINN_arch}
	\vspace{-2mm}
\end{figure*}
For the first time, we apply the approach~\ref{cat5} of using analytical, closed-form gradients for state-space learning of dynamical systems.
We propose a domain-decoupled PINN (DD-PINN) architecture where the time domain is decoupled from the FNN to construct an Ansatz function that enables calculation of gradients in closed form. It is formulated without being limited to a specific Ansatz function, unlike previous work, and can be employed interchangeably with the PINC for physics-loss calculation during training as a state estimator or in control such as for model predictive control.
The DD-PINN is evaluated on three exemplary dynamical systems with varying system complexity and compared to the PINC~\cite{nicodemus_physicsinformed_2022,antonelo_physicsinformed_2024} with regards to its limitations. 

In Sec.~\ref{sec_PINC}, the fundamentals of the PINC are explained and the DD-PINN is formulated in Sec.~\ref{sec_DDPINN}. The validation of the DD-PINN and a comparison to the PINC follows in Sec.~\ref{sec_experiment}, preceeding the conclusion in chapter~\ref{sec_conclusion}.

%% file: figures/ddnn_arch.pdf_tex
\begingroup%
  \makeatletter%
  \providecommand\color[2][]{%
    \errmessage{(Inkscape) Color is used for the text in Inkscape, but the package 'color.sty' is not loaded}%
    \renewcommand\color[2][]{}%
  }%
  \providecommand\transparent[1]{%
    \errmessage{(Inkscape) Transparency is used (non-zero) for the text in Inkscape, but the package 'transparent.sty' is not loaded}%
    \renewcommand\transparent[1]{}%
  }%
  \providecommand\rotatebox[2]{#2}%
  \newcommand*\fsize{\dimexpr\f@size pt\relax}%
  \newcommand*\lineheight[1]{\fontsize{\fsize}{#1\fsize}\selectfont}%
  \ifx\svgwidth\undefined%
    \setlength{\unitlength}{445.03993237bp}%
    \ifx\svgscale\undefined%
      \relax%
    \else%
      \setlength{\unitlength}{\unitlength * \real{\svgscale}}%
    \fi%
  \else%
    \setlength{\unitlength}{\svgwidth}%
  \fi%
  \global\let\svgwidth\undefined%
  \global\let\svgscale\undefined%
  \makeatother%
  \begin{picture}(1,0.64670777)%
    \lineheight{1}%
    \setlength\tabcolsep{0pt}%
    \put(0,0){\includegraphics[width=\unitlength,page=1]{ddnn_arch.pdf}}%
    \put(0.43435138,0.27673781){\makebox(0,0)[lt]{\lineheight{1.25}\smash{\begin{tabular}[t]{l}$\bm{x}_\mathrm{data}$\end{tabular}}}}%
    \put(0.61190532,0.10890821){\makebox(0,0)[lt]{\lineheight{1.25}\smash{\begin{tabular}[t]{l}$\mathcal{L}$\end{tabular}}}}%
    \put(0.45495085,0.12690278){\makebox(0,0)[lt]{\lineheight{1.25}\smash{\begin{tabular}[t]{l}$\mathcal{L}_\mathrm{phys}$\end{tabular}}}}%
    \put(0.45356617,0.19544042){\makebox(0,0)[lt]{\lineheight{1.25}\smash{\begin{tabular}[t]{l}$\mathcal{L}_\mathrm{data}$\end{tabular}}}}%
    \put(0.48490746,0.03366294){\makebox(0,0)[lt]{\lineheight{1.25}\smash{\begin{tabular}[t]{l}Closed-form\end{tabular}}}}%
    \put(0.48490746,0.01426846){\makebox(0,0)[lt]{\lineheight{1.25}\smash{\begin{tabular}[t]{l}gradients (fast training)\end{tabular}}}}%
    \put(0.37611544,0.49643891){\makebox(0,0)[lt]{\lineheight{1.25}\smash{\begin{tabular}[t]{l}AD (slow training)\end{tabular}}}}%
    \put(0.03389961,0.20111144){\makebox(0,0)[lt]{\lineheight{1.25}\smash{\begin{tabular}[t]{l}$\bm{x}_0$\end{tabular}}}}%
    \put(0.03527679,0.14533789){\makebox(0,0)[lt]{\lineheight{1.25}\smash{\begin{tabular}[t]{l}$\bm{u}^\ast_0$\end{tabular}}}}%
    \put(0.04363089,0.09385274){\makebox(0,0)[lt]{\lineheight{1.25}\smash{\begin{tabular}[t]{l}$t$\end{tabular}}}}%
    \put(0.27151246,0.15822623){\makebox(0,0)[lt]{\lineheight{1.25}\smash{\begin{tabular}[t]{l}$\bm{g}$\end{tabular}}}}%
    \put(0.23604101,0.19262537){\makebox(0,0)[lt]{\lineheight{1.25}\smash{\begin{tabular}[t]{l}$\bm{a}$\end{tabular}}}}%
    \put(0.37494213,0.17059904){\makebox(0,0)[lt]{\lineheight{1.25}\smash{\begin{tabular}[t]{l}$\hat{\bm{x}}_t$\end{tabular}}}}%
    \put(0.4460336,0.02361752){\makebox(0,0)[lt]{\lineheight{1.25}\smash{\begin{tabular}[t]{l}$\hat{\dot{\bm{x}}}_t$\end{tabular}}}}%
    \put(0.29769738,0.22117872){\makebox(0,0)[lt]{\lineheight{1.25}\smash{\begin{tabular}[t]{l}$\Delta \hat{\bm{x}}_t$\end{tabular}}}}%
    \put(0.32791077,0.08349535){\makebox(0,0)[lt]{\lineheight{1.25}\smash{\begin{tabular}[t]{l}$\bm{f}_\mathrm{\scriptscriptstyle SSM}$\end{tabular}}}}%
    \put(0.17465095,0.04794193){\makebox(0,0)[lt]{\lineheight{1.25}\smash{\begin{tabular}[t]{l}$\bm{f}_u$\end{tabular}}}}%
    \put(0.40804257,0.08515766){\makebox(0,0)[lt]{\lineheight{1.25}\smash{\begin{tabular}[t]{l}MSE\end{tabular}}}}%
    \put(0.40552067,0.22412749){\makebox(0,0)[lt]{\lineheight{1.25}\smash{\begin{tabular}[t]{l}MSE\end{tabular}}}}%
    \put(0.57930976,0.08648762){\makebox(0,0)[lt]{\lineheight{1.25}\smash{\begin{tabular}[t]{l}$+$\end{tabular}}}}%
    \put(0.91602845,0.10564131){\makebox(0,0)[lt]{\lineheight{1.25}\smash{\begin{tabular}[t]{l}$+$\end{tabular}}}}%
    \put(0.3399784,0.20052988){\makebox(0,0)[lt]{\lineheight{1.25}\smash{\begin{tabular}[t]{l}$+$\end{tabular}}}}%
    \put(0.16772875,0.17167926){\makebox(0,0)[lt]{\lineheight{1.25}\smash{\begin{tabular}[t]{l}FNN\end{tabular}}}}%
    \put(0,0){\includegraphics[width=\unitlength,page=2]{ddnn_arch.pdf}}%
    \put(0.91380208,0.44209459){\makebox(0,0)[lt]{\lineheight{1.25}\smash{\begin{tabular}[t]{l}$\hat{\bm{x}}_{k+1}$\end{tabular}}}}%
    \put(0.94751958,0.08703139){\makebox(0,0)[lt]{\lineheight{1.25}\smash{\begin{tabular}[t]{l}$\hat{\bm{x}}_{k+1}$\end{tabular}}}}%
    \put(0.67141486,0.47154332){\makebox(0,0)[lt]{\lineheight{1.25}\smash{\begin{tabular}[t]{l}$\bm{x}_k$\end{tabular}}}}%
    \put(0.67279204,0.43262219){\makebox(0,0)[lt]{\lineheight{1.25}\smash{\begin{tabular}[t]{l}$\bm{u}_k$\end{tabular}}}}%
    \put(0.68348089,0.39169037){\makebox(0,0)[lt]{\lineheight{1.25}\smash{\begin{tabular}[t]{l}$T$\end{tabular}}}}%
    \put(0.67128747,0.15237592){\makebox(0,0)[lt]{\lineheight{1.25}\smash{\begin{tabular}[t]{l}$\bm{x}_k$\end{tabular}}}}%
    \put(0.80337562,0.23594053){\makebox(0,0)[lt]{\lineheight{1.25}\smash{\begin{tabular}[t]{l}$z^{-1}$\end{tabular}}}}%
    \put(0.80342633,0.55589059){\makebox(0,0)[lt]{\lineheight{1.25}\smash{\begin{tabular}[t]{l}$z^{-1}$\end{tabular}}}}%
    \put(0.67266465,0.11345479){\makebox(0,0)[lt]{\lineheight{1.25}\smash{\begin{tabular}[t]{l}$\bm{u}^\ast_k$\end{tabular}}}}%
    \put(0.68216188,0.06497219){\makebox(0,0)[lt]{\lineheight{1.25}\smash{\begin{tabular}[t]{l}$T$\end{tabular}}}}%
    \put(0.81500835,0.12907948){\makebox(0,0)[lt]{\lineheight{1.25}\smash{\begin{tabular}[t]{l}$\bm{a}$\end{tabular}}}}%
    \put(0.85073607,0.1121134){\makebox(0,0)[lt]{\lineheight{1.25}\smash{\begin{tabular}[t]{l}$\bm{g}$\end{tabular}}}}%
    \put(0.79756959,0.42722744){\makebox(0,0)[lt]{\lineheight{1.25}\smash{\begin{tabular}[t]{l}FNN\end{tabular}}}}%
    \put(0.79362741,0.1202155){\rotatebox{90}{\makebox(0,0)[lt]{\lineheight{1.25}\smash{\begin{tabular}[t]{l}FNN\end{tabular}}}}}%
    \put(0,0){\includegraphics[width=\unitlength,page=3]{ddnn_arch.pdf}}%
    \put(0.44405253,0.44579997){\makebox(0,0)[lt]{\lineheight{1.25}\smash{\begin{tabular}[t]{l}$\mathcal{L}_\mathrm{phys}$\end{tabular}}}}%
    \put(0.42052096,0.56375365){\makebox(0,0)[lt]{\lineheight{1.25}\smash{\begin{tabular}[t]{l}$\mathcal{L}_\mathrm{data}$\end{tabular}}}}%
    \put(0.39068017,0.59664724){\makebox(0,0)[lt]{\lineheight{1.25}\smash{\begin{tabular}[t]{l}$\bm{x}_\mathrm{data}$\end{tabular}}}}%
    \put(0.61120135,0.42973432){\makebox(0,0)[lt]{\lineheight{1.25}\smash{\begin{tabular}[t]{l}$\mathcal{L}$\end{tabular}}}}%
    \put(0.42134594,0.34500586){\makebox(0,0)[lt]{\lineheight{1.25}\smash{\begin{tabular}[t]{l}$\mathcal{L}_\mathrm{IC}$\end{tabular}}}}%
    \put(0.3336176,0.48438764){\makebox(0,0)[lt]{\lineheight{1.25}\smash{\begin{tabular}[t]{l}$\frac{\partial  x}{\partial t}$\end{tabular}}}}%
    \put(0.26314591,0.50579661){\makebox(0,0)[lt]{\lineheight{1.25}\smash{\begin{tabular}[t]{l}$\hat{\bm{x}}_t$\end{tabular}}}}%
    \put(0.03537553,0.53413169){\makebox(0,0)[lt]{\lineheight{1.25}\smash{\begin{tabular}[t]{l}$\bm{x}_0$\end{tabular}}}}%
    \put(0.03615685,0.49604074){\makebox(0,0)[lt]{\lineheight{1.25}\smash{\begin{tabular}[t]{l}$\bm{u}_0$\end{tabular}}}}%
    \put(0.04505837,0.45749206){\makebox(0,0)[lt]{\lineheight{1.25}\smash{\begin{tabular}[t]{l}$t$\end{tabular}}}}%
    \put(0.32572872,0.40105383){\makebox(0,0)[lt]{\lineheight{1.25}\smash{\begin{tabular}[t]{l}$\bm{f}_\mathrm{\scriptscriptstyle SSM}$\end{tabular}}}}%
    \put(0.18139705,0.49108244){\makebox(0,0)[lt]{\lineheight{1.25}\smash{\begin{tabular}[t]{l}FNN\end{tabular}}}}%
    \put(0.35837247,0.54401192){\makebox(0,0)[lt]{\lineheight{1.25}\smash{\begin{tabular}[t]{l}MSE\end{tabular}}}}%
    \put(0.57910034,0.40502934){\makebox(0,0)[lt]{\lineheight{1.25}\smash{\begin{tabular}[t]{l}$+$\end{tabular}}}}%
    \put(0.40851845,0.4042581){\makebox(0,0)[lt]{\lineheight{1.25}\smash{\begin{tabular}[t]{l}MSE\end{tabular}}}}%
    \put(0.26979224,0.32705007){\makebox(0,0)[lt]{\lineheight{1.25}\smash{\begin{tabular}[t]{l}MSE\end{tabular}}}}%
    \put(0.4938904,0.33012617){\makebox(0,0)[lt]{\lineheight{1.25}\smash{\begin{tabular}[t]{l}$\lambda_1$\end{tabular}}}}%
    \put(0.4938904,0.40708909){\makebox(0,0)[lt]{\lineheight{1.25}\smash{\begin{tabular}[t]{l}$\lambda_2$\end{tabular}}}}%
    \put(0.4938904,0.5474865){\makebox(0,0)[lt]{\lineheight{1.25}\smash{\begin{tabular}[t]{l}$\lambda_3$\end{tabular}}}}%
    \put(0.49303373,0.08471269){\makebox(0,0)[lt]{\lineheight{1.25}\smash{\begin{tabular}[t]{l}$\lambda_2$\end{tabular}}}}%
    \put(0.49303373,0.22689779){\makebox(0,0)[lt]{\lineheight{1.25}\smash{\begin{tabular}[t]{l}$\lambda_3$\end{tabular}}}}%
    \put(0.03793986,0.59018026){\makebox(0,0)[lt]{\lineheight{1.25}\smash{\begin{tabular}[t]{l}(a)\end{tabular}}}}%
    \put(0.04279887,0.62580603){\makebox(0,0)[lt]{\lineheight{1.25}\smash{\begin{tabular}[t]{l}Training\end{tabular}}}}%
    \put(0.02418288,0.10737977){\rotatebox{90}{\makebox(0,0)[lt]{\lineheight{1.25}\smash{\begin{tabular}[t]{l}DD-PINN\end{tabular}}}}}%
    \put(0.02391324,0.43219735){\rotatebox{90}{\makebox(0,0)[lt]{\lineheight{1.25}\smash{\begin{tabular}[t]{l}PINC\end{tabular}}}}}%
    \put(0.66709569,0.62518652){\makebox(0,0)[lt]{\lineheight{1.25}\smash{\begin{tabular}[t]{l}Self-loop prediction\end{tabular}}}}%
    \put(0,0){\includegraphics[width=\unitlength,page=4]{ddnn_arch.pdf}}%
    \put(0.66231079,0.59018026){\makebox(0,0)[lt]{\lineheight{1.25}\smash{\begin{tabular}[t]{l}(b)\end{tabular}}}}%
    \put(0.03826343,0.2742086){\makebox(0,0)[lt]{\lineheight{1.25}\smash{\begin{tabular}[t]{l}(c)\end{tabular}}}}%
    \put(0.66264783,0.2742086){\makebox(0,0)[lt]{\lineheight{1.25}\smash{\begin{tabular}[t]{l}(d)\end{tabular}}}}%
  \end{picture}%
\endgroup%

%% file: chapters/chapter2.tex
\section{PRELIMINARIES}
\label{sec_PINC}
In the original PINN for control (PINC)~\cite{antonelo_physicsinformed_2024}, dynamical systems in form of a state-space model (SSM)
\begin{equation}
    \dot{\bm{x}}(t)=\bm{f}_\mathrm{SSM}(\bm{x}(t),\bm{u}(t))
\end{equation}
are approximated using a neural network, where $\bm{x}$ denotes the state vector, $t$ the time domain and $\bm{u}$ an input signal.
The PINC predicts the future state
\begin{equation}
\label{eq_pinc_pred}
\bm{\hat{x}}_t =\bm{f}_\mathrm{NN}(\bm{x}_0,\bm{u}_0,t,\bm{\theta}) \approx \bm{x}(t)
\end{equation}
with the feed-forward neural network $\bm{f}_\mathrm{NN}$ continuously over the sampling interval
\begin{equation}
\label{eq_sample_time_int}
0 \leq t \leq T_\mathrm{s}
\end{equation}
in which the excitation is assumed constant and subject to boundaries
\begin{equation}
\label{eq_sample_int}
\begin{split}
\bm{x}_\mathrm{min} \leq \bm{x}_0 \leq \bm{x}_\mathrm{max}, \\
\bm{u}_\mathrm{min} \leq \bm{u}_0 \leq \bm{u}_\mathrm{max}. \\
\end{split}
\end{equation}

The parameters, i.e., the weights and biases $\bm{\theta}$ of the FNN are trained according to the process visualized in Fig.~\ref{fig_DDPINN_arch}(a) using a custom physics-informed loss function 
\begin{equation}
    \mathcal{L}(\bm{\theta})=\lambda_1 \mathcal{L}_\mathrm{IC}(\bm{\theta})+\lambda_2 \mathcal{L}_\mathrm{phys}(\bm{\theta})+\lambda_3 \mathcal{L}_\mathrm{data}(\bm{\theta})
    \label{eq_loss_function}
\end{equation}
consisting of the initial-condition loss 
\begin{equation}
\mathcal{L}_\mathrm{IC}=\mathrm{MSE}(\bm{x}_0,\bm{\hat{x}}_0),   
\end{equation}
physics loss
\begin{equation}
\mathcal{L}_\mathrm{phys}=\mathrm{MSE}\Big( \frac{\partial}{\partial t}\bm{\hat{x}}_t ,\bm{f}_\mathrm{SSM}\big(\bm{\hat{x}}_t,\bm{u}_0\big)\Big),  
\end{equation}
and data loss 
\begin{equation}
\mathcal{L}_\mathrm{data}=\mathrm{MSE}(\bm{\hat{x}}_t,\bm{x}_\mathrm{data}).
\end{equation}
The mean squared error is denoted by $\mathrm{MSE}()$.
The initial-condition loss, if sufficiently minimized, ensures that for $t=0$ the PINC predicts the initial state $\bm{x}_0$. The physics loss $\mathcal{L}_\mathrm{phys}$ includes the system's dynamics equation in state-space form and therefore governs learning of the physics model. The derivative of the predicted state $\frac{\partial}{\partial t}\bm{\hat{x}}_t$ is calculated through graph-based automatic differentiation with respect to the time input $t$. In principle, these two losses are sufficient to learn the system dynamics. Optionally, a data loss $\mathcal{L}_\mathrm{data}$ can be added, to include data sets obtained from real hardware to incorporate effects that may not be captured by the first-principles model.
As these losses generally have different magnitudes and convergence behaviors, they are weighted using the factors $\lambda_i$, which are either determined empirically or by loss-weighting schemes as mentioned in Sec.~\ref{sec_intro}.

In the training process, the loss functions $\mathcal{L}_\mathrm{IC}$ and $\mathcal{L}_\mathrm{phys}$ are evaluated on randomly sampled collocation points in the intervals given in~\eqref{eq_sample_time_int} and~\eqref{eq_sample_int}. One sampling method commonly applied is Latin hypercube sampling. Using backpropagation, the FNN parameters $\bm{\theta}$ are optimized using the total loss in~\eqref{eq_loss_function}.

After training, the PINC can be employed for dynamic state prediction in self loop as visualized in Fig.~\ref{fig_DDPINN_arch}(b) where $k$ indicates the current step. The step size
$T<T_\mathrm{s}$ is chosen constant
for numerical integration at an operation frequency $f=\frac{1}{T}$.
\section{DOMAIN-DECOUPLED PINN}
\label{sec_DDPINN}
We propose an alternative architecture to the PINC, called the domain-decoupled physics-informed neural network (DD-PINN) to address the limitations listed in Sec.~\ref{sec_intro}. The architecture of the DD-PINN in comparison to the PINC is visualized in Fig.~\ref{fig_DDPINN_arch}(c--d) for the training and prediction process, respectively. The main modification involves decoupling the time domain via
\begin{align}
    \label{eq_ddnn_pred}
    \bm{\hat{x}}_t & =\bm{g}  \big( \bm{f}_\mathrm{NN}(\bm{x}_0,\bm{u}_0,\bm{\theta}),t \big)+\bm{x}_0 \notag \\
    & = \bm{g}(\bm{a},t) +\bm{x}_0
\end{align}
where $\bm{f}_\mathrm{NN}$ now predicts the vector
\begin{equation}
\bm{a} = \bm{f}_\mathrm{NN}(\bm{x}_0,\bm{u}_0,\bm{\theta})
\end{equation}
which is used in an Ansatz function $\bm{g}$. We restrict $\bm{g}$ to the following two properties:
\begin{enumerate}
    \item $\bm{g}$\ is differentiable in closed form $\forall t\in [0,T_\mathrm{s}]$ 
    \item $\bm{g}(\bm{a},0) \equiv \bm{0}$
\end{enumerate}
The first property allows us to formulate the derivative of the predicted state as
\begin{equation}
    \frac{\partial}{\partial t}\bm{\hat{x}}_t = \dot{\bm{g}}  \big( \bm{a},t \big).
\end{equation}
Therefore, we can calculate the gradients of the predicted state with respect to the time domain in analytical, closed form and do not require computationally expensive automatic differentiation, addressing limitation~\ref{lim1}.
Additionally, the initial condition is always fulfilled, i.e., 
\begin{equation}
\mathcal{L}_\mathrm{IC}\equiv 0.
\end{equation}
This addresses limitation~\ref{lim3} due to the second property of $\bm{g}$ and introducing the initial state $\bm{x}_0$ in~\eqref{eq_ddnn_pred}. The latter is comparable to a residual neural network (ResNet) architecture \cite{he_deep_2016}.
The loss-balancing problem arising from~\eqref{eq_loss_function} is then drastically simplified. If no real data is used for training and only a fast surrogate model is to be learned, no loss balancing is even necessary, as only one loss exists.
The consideration of lim.~\ref{lim2} can be found in section~\ref{subsec_uext}.
\subsection{Formulating the Ansatz Function}
The Ansatz function
\begin{align}
    \bm{g} & = (g_1,\ldots, g_j, \ldots, g_m)^\top
\end{align}
consists of $m$ residual predictions $g_j$, where $m$ is the number of states in the state-space model and $g_j$ the residual prediction of the $j$-th state variable $\Delta \hat{x}_{t,j}$ for a given $t$. The residual state prediction is constructed using $n_g$ sub-functions.
\begin{equation}
    g_j=\sum_{i=1}^{n_g} \alpha_{ij} \big( \phi_g(\beta_{ij}t+\gamma_{ij})-\phi_g(\gamma_{ij}) \big)
    \label{eq_gj}
\end{equation}
where $\phi_g$ is a differentiable activation or base construction function. We observe that the subtrahend in~\eqref{eq_gj} ensures that $g_j(t=0)\equiv 0$, and the total length of the vector
\begin{equation}
\bm{a}=(\bm{\alpha}^\top,\bm{\beta}^\top,\bm{\gamma}^\top)^\top
\end{equation}
with coefficient vectors $\bm{\alpha},\bm{\beta},\bm{\gamma}$
results to $3 \cdot m \cdot n_g$.
Finally, we obtain the derivative of the predicted state 
\begin{equation}
    \dot{g}_j=\sum_{i=1}^{n_g} \alpha_{ij} \beta_{ij}  \dot{\phi}_g(\beta_{ij}t+\gamma_{ij}) 
    \label{eq_dgj}
\end{equation}
using the chain rule. Alternatively, a damping term can be added to increase the numbers of parameters in the sub-functions
\begin{align}
    g_{\mathrm{d}j} &= \sum_{i=1}^{n_g} \alpha_{ij} \big(\mathrm{e}^{-\delta_{ij}t} \phi_g(\beta_{ij}t+\gamma_{ij}) -\phi_g(\gamma_{ij}) \big), \notag \\
    \dot{g}_{\mathrm{d}j} &=\sum_{i=1}^{n_g} \alpha_{ij} \mathrm{e}^{-\delta_{ij}t} \big( \beta_{ij}\dot{\phi}_g(\beta_{ij}t+\gamma_{ij}) \notag \\& \quad \quad \quad -\delta_{ij}\phi_g(\beta_{ij}t+\gamma_{ij}) \big),
    \label{eq_gj_damp}
\end{align}
 where now the length of vector $\bm{a}=(\bm{\alpha}^\top,\bm{\beta}^\top,\bm{\gamma}^\top,\bm{\delta}^\top)^\top$ results to $4 \cdot m \cdot n_g$.

The inclusion of the alternative damping term and the choice of the base construction function $\phi_g$ may be decided based on the given state-space model characteristics or empirical testing, which is similar to the choice of a suitable activation functions of neural networks. In this study,
\begin{equation}
\phi_g(x)=\sin(x) \quad \text{and} \quad
 \dot{\phi}_g(x)=\cos(x)
\end{equation}
is used, which achieves good results on all three different systems in Sec.~\ref{sec_experiment}. However, any other differentiable function may be used.

\subsection{Higher-order Excitation}
\label{subsec_uext}
In order to address limitation~\ref{lim2} concerning the zero-order hold assumption, we introduce the option for higher-order excitation input to the DD-PINN. As visualized in Fig.~\ref{fig_DDPINN_arch}(c), the input to the FNN is the extended excitation vector $\bm{u}^\ast_0$, that contains a concatenation of multiple excitation vectors inside the interval $[0,T]$. The function $\bm{f}_u$ calculates the interpolated excitation vector
\begin{equation}
\label{eq_uext}
    \bm{u}_t=\bm{f}_u(\bm{u}^\ast_0 ,t)
\end{equation}
fed into the dynamics equation $\bm{f}_\mathrm{SSM}$.
In this study, we limit the analysis to zero-, first- and second-order excitation, hereinafter referred to as the degree $\delta_u$ of $\bm{u}$, using the following polynomial interpolations with $t_\ast=\frac{t}{T}$:
 \begin{enumerate}
     \item Zero-order hold, $\delta_u=0$, similar to PINC:
     \begin{equation}
         \bm{u}^\ast_0=\bm{u}_0, \quad \bm{f}_u=\bm{u}_0
     \end{equation}
     \item First order, $\delta_u=1$:
    \begin{equation}
         \bm{u}^\ast_0=(\bm{u}_0^\top,\bm{u}_{T}^\top)^\top, \quad \bm{f}_u=\bm{u}_T t_\ast + (1-t_\ast)\bm{u}_0
     \end{equation}
     \item Second order, $\delta_u=2$: 
    \begin{align}
         \bm{u}^\ast_0 & =(\bm{u}_0^\top,\bm{u}_{T/2}^\top,\bm{u}_{T}^\top)^\top,\notag \\ \quad \bm{f}_u & =
         (2\bm{u}_0 -4\bm{u}_{T/2} +2\bm{u}_{T}){t_\ast}^2 \notag \\ & \quad + (-3\bm{u}_0 +4\bm{u}_{T/2} -1\bm{u}_{T}){t_\ast}+\bm{u}_0.
    \end{align}
 \end{enumerate}
Here, $\bm{u}_{T/2}=\bm{u}(t=\frac{T}{2})$ and $\bm{u}_{T}=\bm{u}(t=T)$ denote the excitation at half and full step interval.
The FNN then implicitly learns the polynomial interpolation of the components in $\bm{u}^\ast_0$. It can be noted that instead of this polynomial interpolation, a polynomial Taylor expansion may also be used. If applied for model predictive control, a DD-PINN trained for first-order or second-order input can still predict the system behavior under zero-order-hold assumption with $\bm{u}_0 = \bm{u}_{T/2} = \bm{u}_{T}$. Depending on the system‘s characteristics, an optimized excitation vector for the (large) time steps within MPC could be transformed into a first/second-order excitation to better approximate fine changes within small time steps outside the MPC loop.

\subsection{Application of the DD-PINN to Control Scenarios and its Generality}
When the PINC is operated in self-loop mode for prediction of the dynamical system behavior as visualized in Fig.~\ref{fig_DDPINN_arch}(b), only a relatively short time horizon is predicted at once compared to applications of the classical PINN.
We therefore argue it is sufficiently accurate to model this short interval with the Ansatz function $g$ of the DD-PINN without the need of a high number of construction functions $n_g$.
The DD-PINN also maintains the generality of the PINC, as the parameter $n_g$ as well as the FNN size can be chosen sufficiently large to model any arbitrary, continuous function.
It shall also be noted, that the concept proposed here is not limited to the time domain but may be applied to the spatial domain or others for the solution of ODEs/PDEs.

%% file: chapters/chapter3.tex
\section{VALIDATION}
\begin{figure*}[ht]
	\centering
	\resizebox{1\linewidth}{!}{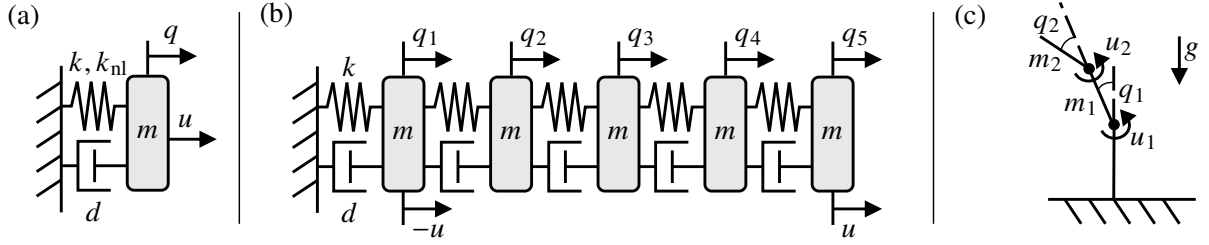}
	\caption{Three dynamical systems for evaluation: (a) nonlinear mass-spring-damper system, (b) five-mass-chain system, (c) two-link manipulator system used in~\cite{nicodemus_physicsinformed_2022}.}
	\label{fig_dyn_systems}
	\vspace{-2mm}
\end{figure*} 
\label{sec_experiment}
In this section, the PINC and the DD-PINN are evaluated on three dynamical systems visualized in Fig.~\ref{fig_dyn_systems} for comparison.
The PINC and DD-PINN are both trained and implemented using \mbox{PyTorch}. LRA is used to balance the initial-condition loss (only PINC), the data loss, and the physics loss. A data loss is only included for the nonlinear mass-spring-damper system. It is noted that these data points are not necessary for training but demonstrate the capability of the PINC and DD-PINN to include data that may be obtained from a real system.
For all systems, the Adam optimization algorithm is used with an initial learning rate of $\alpha_\mathrm{init}=0.001$.
A learning-rate scheduler that reduces the learning rate on a plateau is employed and the training is stopped early if the rate falls below a certain threshold of $\alpha_\mathrm{min}=5\cdot 10^{-8}$. For the FNN in all models, the Gaussian error linear unit (GELU) activation function is used~\cite{hendrycks_gaussian_2016}. The first system is trained on an Intel Cascade Lake Xeon Gold 6230N CPU with 16 GB of RAM and the latter two on an Intel Core i9-10900X CPU with 64 GB of RAM. The training and neural network parameters for each system are listed in Table~\ref{tab_net_param} in the appendix and an overview of the results are given in Table~\ref{tab_results_msd}--\ref{tab_results_other}.

\subsection{Nonlinear Mass-spring-damper System}
\label{subsec_nlsmd_experiment}
The nonlinear mass-spring-damper system visualized in Fig.~\ref{fig_dyn_systems}(a) consists of a mass $m$ that is connected to a fixed wall through a spring with linear stiffness $k$ and a cubic component of $k_\mathrm{nl}$ as well as a damper with coefficient $d$. An external force $u$ is applied to the mass and the displacement $q$ and velocity $\dot{q}$ denote the system's state $\bm{x}=(q,\dot{q})^\top$.
The state-space model results to
\begin{equation}
    \label{eq_ssm_nlsmd}
\dot{\bm{x}}=\begin{pmatrix}
\dot{q} \\
\frac{u-d \dot{q} - k q - k_\mathrm{nl} {q}^3 }{m}
\end{pmatrix}
\end{equation}
with $m=\qty{0.001}{kg}$, $d=\qty{0.001}{Ns/m}$, $k=\qty{1}{N/m}$, and $k_\mathrm{nl}=\qty{15}{N/m^3}$.

Both architectures, the PINC and the DD-PINN, are evaluated for three operation frequencies $f\in \{\qty{50}{Hz},\qty{100}{Hz},\qty{200}{Hz}\}$. The DD-PINN is further evaluated in all combinations for zero-, first-, and second-order excitation $\delta_u \in \{0, 1, 2\}$ as described in section~\ref{subsec_uext} and for different numbers of base construction functions $n_g \in \{5, 20, 50\}$ -- resulting in a total of 27 DD-PINN models trained. For those models, a  trigonometric base function $\phi_g(x)=\sin(x)$ is used including the optional damping factor described in~\eqref{eq_gj_damp}.
\begin{figure}[t]
\begin{center}
\includegraphics[width=\columnwidth]{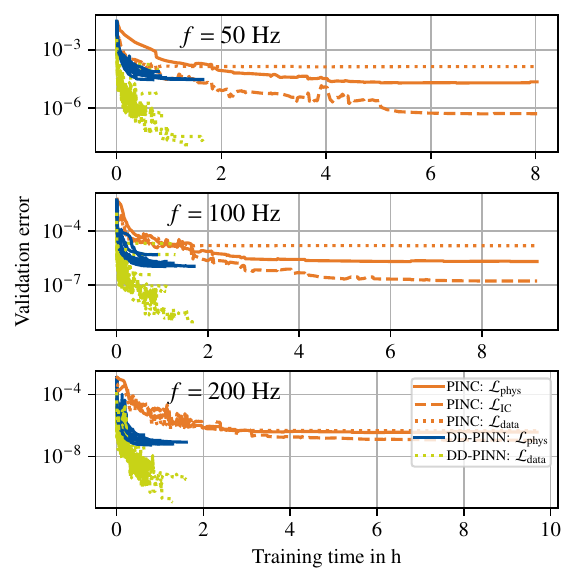}
\caption{Nonlinear mass-spring-damper system learning; Reduction of validation errors over time. The DD-PINN training time is significantly lower.}
\label{fig_nlsmd_main_a}
\end{center}
\end{figure}
\begin{figure}[t]
\begin{center}
\includegraphics[width=\columnwidth]{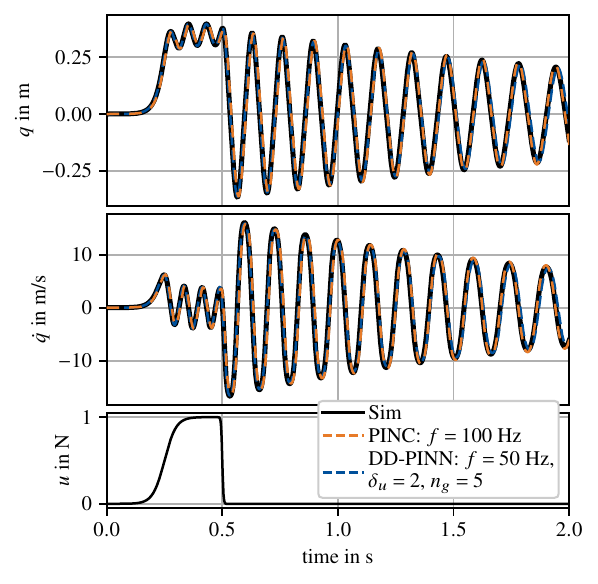}
\caption{Nonlinear mass-spring-damper system self-loop prediction of a test path for selected PINN and DD-PINN models. Both models exhibit high prediction accuracy.}
\label{fig_nlsmd_main_b}
\end{center}
\end{figure}
\begin{figure}[t]
\begin{center}
\includegraphics[width=\columnwidth]{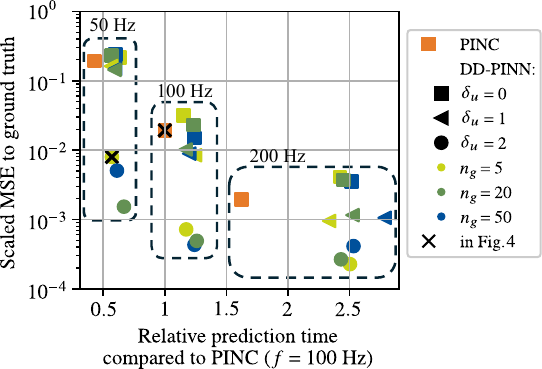}
\caption{Parameter study of the nonlinear mass-spring-damper system learning results; Scaled MSE to ground truth of $1$ s self-loop prediction path over relative prediction time. DD-PINN models achieve significantly higher prediction accuracy for slightly slower prediction time.}
\label{fig_nlsmd_comparison}
\end{center}
\end{figure}

For both architectures, an FNN with $96$ neurons and one hidden layer is chosen.
The PINC is trained for a maximum of \num{4000} epochs on \num{20000} collocation points, \num{40000} initial condition points and \num{2000} data points distributed over $400$ batches. The DD-PINN is trained for a maximum of \num{4000} epochs as well. The fast calculation of closed-form gradients allows us to increase the count of collocation points to \num{375000} distributed over \num{1500} batches.
Data points are obtained from a simulation of the state-space model~\eqref{eq_ssm_nlsmd}.

The sampling ranges for the collocation points
\begin{align}
\label{eq_sample_nlsmd}
&(q_\mathrm{min},\dot{q}_\mathrm{min},u_\mathrm{min},t_\mathrm{min})\notag \\
\quad &=(\qty{-0.4}{m},\qty{-18}{m/s},\qty{-1}{N},\qty{0}{s}), \notag  \\
&(q_\mathrm{max},\dot{q}_\mathrm{max},u_\mathrm{max},t_\mathrm{max})\notag \\
\quad &=(\qty{0.4}{m},\qty{18}{m/s},\qty{1}{N},\frac{1.1}{f})
\end{align}
are also used for normalization of the network inputs.

The training process is visualized in Fig.~\ref{fig_nlsmd_main_a} showing the validation physics loss $\mathcal{L}_\mathrm{phys}$ decrease over time. We observe that the DD-PINN models converge in average $8.7$ times (minimum $4.7$, maximum $18.9$) faster than the PINC models despite a $9.375$ times higher count of collocation points per epoch. The minimum validation error reached by the DD-PINN networks is also lower in the case of 100 and 200 Hz, but slightly higher for 50 Hz. It can also be seen that the DD-PINN reaches a drastically lower validation error for the data loss.
The trained systems are evaluated on a test path in self-loop prediction of which the PINC model at $f=100$ Hz and a selected DD-PINN model ($f=\qty{50}{Hz}$, $\delta_u=2$, $n_g=5$) are visualized in Fig.~\ref{fig_nlsmd_main_b}. We observe an accurate and stable self-loop prediction of both models on a time horizon of more than $5$ s.

A comparison of accuracy over self-loop prediction time is given in Fig.~\ref{fig_nlsmd_comparison}. Here, the MSE between the predicted path from Fig.~\ref{fig_nlsmd_main_b} and the ground truth is calculated until $t=1$ s and visualized over the prediction time in relation to the PINC model at $f=100$ Hz. We observe that a higher frequency, i.e., a shorter shooting interval generally corresponds to higher prediction accuracy, coming at cost of prediction time. Looking at each group of frequencies 50, 100, and $\qty{200}{Hz}$ respectively, the DD-PINN achieves up to $2.1$, $1.7$, and $0.9$ magnitudes higher accuracy with $\delta_u>0$. The prediction time is only slightly higher at approx. $1.2$ times for the $100$ Hz models and approx. $1.5$ times for the other two groups. A more detailed overview of the results is also given in the appendix in Table~\ref{tab_results_msd}.

\subsection{Five-mass-chain System}
The second system is a one-dimensional coupled chain of five masses as visualized in Fig.~\ref{fig_dyn_systems}(b). They are connected at one side to a fixed wall and between each other with a total of five stiffnesses $k$ and dampers $d$. As an external input, the same load $u$ is applied antagonistically at the first and last mass. The state vector is defined in relative coordinates with displacements and velocities
 \begin{align}
\bm{q}=(q_1,q_2,q_3,q_4,q_5)^\top, \notag \\
\dot{\bm{q}}=(\dot{q}_1,\dot{q}_2,\dot{q}_3,\dot{q}_4,\dot{q}_5)^\top,
\end{align}
 resulting in a total of ten states in the state vector
 \begin{equation}
 \label{eq_1dm_state_vector}
\bm{x}=(\bm{q}^\top,\dot{\bm{q}}^\top)^\top.
\end{equation}
The state vector is evaluated on a physics loss function based on the equation of motion
\begin{equation}
\label{eq_1dm_eqm}
    \bm{M} \ddot{\bm{q}}(t) + \bm{D} \dot{\bm{q}}(t) + \bm{K} \bm{q}(t) = \bm{P} u(t)
\end{equation}
with mass matrix $\bm{M}$, stiffness matrix $\bm{K}$, damping Matrix $\bm{D}$, and vector
\begin{equation}
    \bm{P}=(-1,0,0,0,1)^\top.
\end{equation}

Two DD-PINN models are trained at $f=50$ Hz with $\delta_u \in \{0,2\}$. The hyper parameters are set to $n_g = 20$, $\phi_g(x)=\sin(x)$ and the optional damping factor given in~\eqref{eq_gj_damp} is used.
This system is hard to predict accurately under the ZOH assumption showcased by one PINC model trained at $f=\qty{50}{Hz}$.
For the DD-PINN and PINC, an FNN with $128$ neurons and two hidden layers is chosen.
The PINC is trained on \num{20000} collocation points, \num{40000} initial condition points over $400$ batches and without data loss.
The DD-PINN is trained for \num{2500} epochs, using \num{250000} collocation points distributed over $500$ batches.
The sampling ranges for the collocation points and input normalization are determined based on the simulated test path used, including a margin of $10\%$.
\begin{figure}[t]
\begin{center}
\includegraphics[width=\columnwidth]{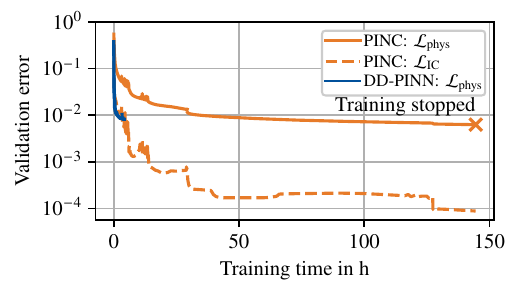}
\caption{Five-mass-chain system learning: Reduction of validation errors over training time. DD-PINN converged within \qty{5}{h}, while the PINC training was stopped after \qty{145}{h}.}
\label{fig_1dm_learning}
\end{center}
\end{figure}
\begin{figure}[t]
\begin{center}
\includegraphics[width=\columnwidth]{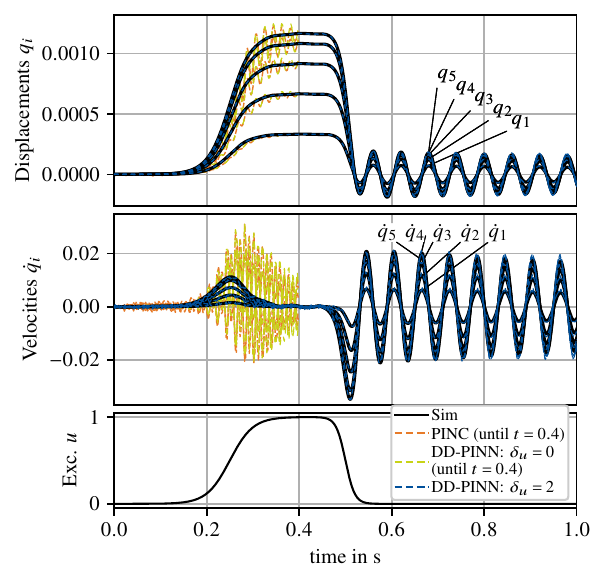}
\caption{DD-PINN self-loop prediction of a test path for the five-mass-chain system including poor predictions for models with ZOH assumption until $t=\qty{0.4}{s}$. The DD-PINN's higher order excitation enables smooth prediction.}
\label{fig_1dm_prediction}
\end{center}
\end{figure}

Fig.~\ref{fig_1dm_learning} shows the validation physics loss reduction during the training process. The DD-PINNs both converged at around \qty{4}{h} while the PINC training was stopped after \qty{145}{h} and \num{2400} epochs of training.
As we can observe in Fig.~\ref{fig_1dm_prediction}, the ZOH assumption is not sufficent for a prediction frequency at 50 Hz as strong oscillations are visible in both, the PINC's and the DD-PINN's self-loop prediction (only plotted until $t=\qty{0.4}{s}$). This is due to the interval-wise constant input acting as a step excitation that excites higher modes of the system.

It can be seen that having an interval-wise quadratic excitation input to the DD-PINN, the predicted path becomes stable and accurate to the simulated system response.
The self-loop prediction time for the DD-PINN models was measured to be $33\%$ and $34\%$ slower than the PINC for $\delta_u=0$ and $\delta_u=2$, respectively, noting that the increased order in excitation does not come along with an increase in prediction time for the DD-PINN.

\subsection{Two-link Manipulator System}
The two-link manipulator system visualized in Fig.~\ref{fig_dyn_systems}(c) denotes a Schunk PowerCube robot with dynamics identified in~\cite{fehr_identification_2022}. In previous work~\cite{nicodemus_physicsinformed_2022}, a PINC was used in MPC for fast and accurate prediction of the system dynamics. The manipulator consists of one fixed link in upright position following two sequential links connected by two joints at angle $q_1$ and $q_2$ at which the motor currents $u_1$, $u_2$ apply. With the generalized coordinates $\bm{q}=(q_1,q_2)^\top$ and the state vector $\bm{x}=(\bm{q}^\top,\dot{\bm{q}}^\top)^\top$,
the state-space model results to
\begin{equation}
    \label{eq_ssm_nc}
\dot{\bm{x}}=\begin{pmatrix}
\dot{\bm{q}} \\
\bm{M}(\bm{q})^{-1} \big(  \bm{h}(\bm{q},\dot{\bm{q}})  - \bm{k}(\bm{q},\dot{\bm{q}}) + \bm{B} \bm{u}      \big)
\end{pmatrix}
\end{equation}
with mass matrix $\bm{M}$, input matrix $\bm{B}$ (including motor constant), vector of centrifugal and Coriolis forces $\bm{k}$ and vector of gravitational and damping forces $\bm{h}$ adopted from~\cite{nicodemus_physicsinformed_2022}.

\begin{figure*}[t]
\begin{center}
\includegraphics[width=\linewidth]{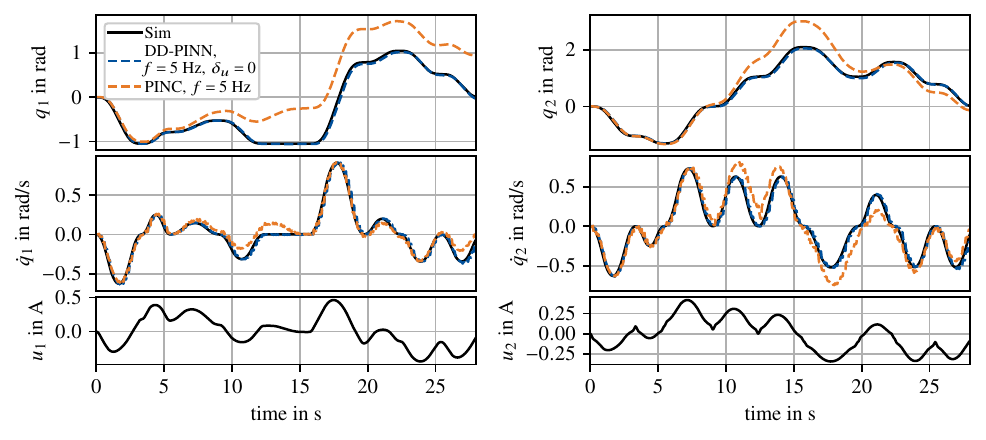}
\caption{PINC and DD-PINN self-loop prediction of a test path for the two-link manipulator system. The DD-PINN's prediction is stable while the PINC's prediction diverges.}
\label{fig_nc_prediction}
\end{center}
\end{figure*}
\begin{figure}[t]
\begin{center}
\includegraphics[width=\columnwidth]{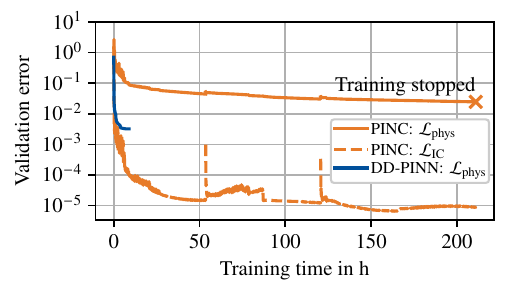}
\caption{Two-link manipulator system learning: Reduction of validation errors over training time.  DD-PINN converged within \qty{7.4}{h}, while the PINC training was stopped after more than \qty{210}{h}.}
\label{fig_nc_learning}
\end{center}
\end{figure}
One PINC and one DD-PINN model at $f=5$ Hz are trained. The DD-PINN is set to $\delta_u=0$ for direct comparison, using $n_g = 20$, $\phi_g(x)=\sin(x)$ and the optional damping factor given in~\eqref{eq_gj_damp}.
For the PINC, an FNN with $64$ neurons and four hidden layers (equivalent to~\cite{nicodemus_physicsinformed_2022}) is chosen and $128$ neurons with two hidden layers for the DD-PINN.
The PINC is trained for \num{7000} epochs on \num{20000} collocation points, \num{40000} initial condition points over $400$ batches and without data loss.
The DD-PINN is trained for \num{2500} epochs, using an increased count of collocation points of $1 \cdot 10^6$ distributed over \num{1000} batches. 
The used sampling ranges for the collocation points and input normalization are
\begin{align}
\label{eq_sample_nc}
&(q_{1,\mathrm{min}}, q_{2,\mathrm{min}}, \dot{q}_{1,\mathrm{min}}, \dot{q}_{2,\mathrm{min}}, u_{1,\mathrm{min}}, u_{2,\mathrm{min}}, t_\mathrm{min}) \notag \\
&= (\SI{-\pi}{}, \qty{-\pi}{}, \qty{-1}{s^{-1}}, \qty{-1}{s^{-1}}, \qty{-0.6}{A}, \qty{-0.6}{A}, \qty{0}{s}), \notag \\
&(q_{1,\mathrm{max}}, q_{2,\mathrm{max}}, \dot{q}_{1,\mathrm{max}}, \dot{q}_{2,\mathrm{max}}, u_{1,\mathrm{max}}, u_{2,\mathrm{max}}, t_\mathrm{max}) \notag \\
&= (\qty{\pi}{}, \qty{\pi}{}, \qty{1}{s^{-1}}, \qty{1}{s^{-1}}, \qty{0.6}{A}, \qty{0.6}{A},  \frac{1.1}{f}).
\end{align}

Fig.~\ref{fig_nc_learning} shows the validation physics loss reduction during the training process. The training of the PINC did not converge within $7000$ epochs  and was discontinued due to the exceedingly high training time of over 210 h. The DD-PINN converged in a comparatively short time of 7.4 h while reaching a significantly lower validation error.
This difference shows effect in the self-loop prediction on a custom test path in Fig.~\ref{fig_nc_prediction}. Here, the PINC's prediction diverges after \qty{7}{s}. This is comparable to previous work where the PINC prediction diverged in self-loop after approx. \qty{3}{s} for the same system~\cite{nicodemus_physicsinformed_2022}. In contrast, the prediction of the DD-PINN remains stable and accurate for the whole \qty{28}{s}-long test path. For this system, the self-loop prediction time for the DD-PINN model was measured to be approximately equal to the computation time of the PINC. The results of the two-link manipulator system as well as the five-mass-chain system are also given in the appendix in Table~\ref{tab_results_other}.

%% file: figures/dynsys.pdf_tex
\begingroup%
  \makeatletter%
  \providecommand\color[2][]{%
    \errmessage{(Inkscape) Color is used for the text in Inkscape, but the package 'color.sty' is not loaded}%
    \renewcommand\color[2][]{}%
  }%
  \providecommand\transparent[1]{%
    \errmessage{(Inkscape) Transparency is used (non-zero) for the text in Inkscape, but the package 'transparent.sty' is not loaded}%
    \renewcommand\transparent[1]{}%
  }%
  \providecommand\rotatebox[2]{#2}%
  \newcommand*\fsize{\dimexpr\f@size pt\relax}%
  \newcommand*\lineheight[1]{\fontsize{\fsize}{#1\fsize}\selectfont}%
  \ifx\svgwidth\undefined%
    \setlength{\unitlength}{415.09924701bp}%
    \ifx\svgscale\undefined%
      \relax%
    \else%
      \setlength{\unitlength}{\unitlength * \real{\svgscale}}%
    \fi%
  \else%
    \setlength{\unitlength}{\svgwidth}%
  \fi%
  \global\let\svgwidth\undefined%
  \global\let\svgscale\undefined%
  \makeatother%
  \begin{picture}(1,0.19803871)%
    \lineheight{1}%
    \setlength\tabcolsep{0pt}%
    \put(0,0){\includegraphics[width=\unitlength,page=1]{dynsys.pdf}}%
    \put(-0.00079047,0.17925574){\makebox(0,0)[lt]{\lineheight{1.25}\smash{\begin{tabular}[t]{l}(a)\end{tabular}}}}%
    \put(0.20908716,0.18130613){\makebox(0,0)[lt]{\lineheight{1.25}\smash{\begin{tabular}[t]{l}(b)\end{tabular}}}}%
    \put(0.78747748,0.17925574){\makebox(0,0)[lt]{\lineheight{1.25}\smash{\begin{tabular}[t]{l}(c)\end{tabular}}}}%
    \put(0.04943237,0.14007146){\makebox(0,0)[lt]{\lineheight{1.25}\smash{\begin{tabular}[t]{l}$k,k_\mathrm{nl}$\end{tabular}}}}%
    \put(0,0){\includegraphics[width=\unitlength,page=2]{dynsys.pdf}}%
    \put(0.28359624,0.01374863){\makebox(0,0)[t]{\lineheight{1.25}\smash{\begin{tabular}[t]{c}$d$\end{tabular}}}}%
    \put(0.32899203,0.08316367){\makebox(0,0)[t]{\lineheight{1.25}\smash{\begin{tabular}[t]{c}$m$\end{tabular}}}}%
    \put(0,0){\includegraphics[width=\unitlength,page=3]{dynsys.pdf}}%
    \put(0.28387696,0.13548156){\makebox(0,0)[t]{\lineheight{1.25}\smash{\begin{tabular}[t]{c}$k$\end{tabular}}}}%
    \put(0,0){\includegraphics[width=\unitlength,page=4]{dynsys.pdf}}%
    \put(0.0706512,0.01449241){\makebox(0,0)[t]{\lineheight{1.25}\smash{\begin{tabular}[t]{c}$d$\end{tabular}}}}%
    \put(0.11604697,0.08390745){\makebox(0,0)[t]{\lineheight{1.25}\smash{\begin{tabular}[t]{c}$m$\end{tabular}}}}%
    \put(0,0){\includegraphics[width=\unitlength,page=5]{dynsys.pdf}}%
    \put(0.14129515,0.09286217){\makebox(0,0)[lt]{\lineheight{1.25}\smash{\begin{tabular}[t]{l}$u$\end{tabular}}}}%
    \put(0.12897056,0.16752894){\makebox(0,0)[lt]{\lineheight{1.25}\smash{\begin{tabular}[t]{l}$q$\end{tabular}}}}%
    \put(0.33604506,0.1650681){\makebox(0,0)[lt]{\lineheight{1.25}\smash{\begin{tabular}[t]{l}$q_1$\end{tabular}}}}%
    \put(0.33580535,0.00110101){\makebox(0,0)[lt]{\lineheight{1.25}\smash{\begin{tabular}[t]{l}$-u$\end{tabular}}}}%
    \put(0,0){\includegraphics[width=\unitlength,page=6]{dynsys.pdf}}%
    \put(0.69360596,0.00110101){\makebox(0,0)[lt]{\lineheight{1.25}\smash{\begin{tabular}[t]{l}$u$\end{tabular}}}}%
    \put(0,0){\includegraphics[width=\unitlength,page=7]{dynsys.pdf}}%
    \put(0.41859678,0.08316367){\makebox(0,0)[t]{\lineheight{1.25}\smash{\begin{tabular}[t]{c}$m$\end{tabular}}}}%
    \put(0,0){\includegraphics[width=\unitlength,page=8]{dynsys.pdf}}%
    \put(0.42587602,0.16506805){\makebox(0,0)[lt]{\lineheight{1.25}\smash{\begin{tabular}[t]{l}$q_2$\end{tabular}}}}%
    \put(0,0){\includegraphics[width=\unitlength,page=9]{dynsys.pdf}}%
    \put(0.50803324,0.08316367){\makebox(0,0)[t]{\lineheight{1.25}\smash{\begin{tabular}[t]{c}$m$\end{tabular}}}}%
    \put(0,0){\includegraphics[width=\unitlength,page=10]{dynsys.pdf}}%
    \put(0.51531249,0.16506805){\makebox(0,0)[lt]{\lineheight{1.25}\smash{\begin{tabular}[t]{l}$q_3$\end{tabular}}}}%
    \put(0,0){\includegraphics[width=\unitlength,page=11]{dynsys.pdf}}%
    \put(0.597018,0.08316367){\makebox(0,0)[t]{\lineheight{1.25}\smash{\begin{tabular}[t]{c}$m$\end{tabular}}}}%
    \put(0,0){\includegraphics[width=\unitlength,page=12]{dynsys.pdf}}%
    \put(0.60429719,0.16506805){\makebox(0,0)[lt]{\lineheight{1.25}\smash{\begin{tabular}[t]{l}$q_4$\end{tabular}}}}%
    \put(0,0){\includegraphics[width=\unitlength,page=13]{dynsys.pdf}}%
    \put(0.68629009,0.08316367){\makebox(0,0)[t]{\lineheight{1.25}\smash{\begin{tabular}[t]{c}$m$\end{tabular}}}}%
    \put(0,0){\includegraphics[width=\unitlength,page=14]{dynsys.pdf}}%
    \put(0.69356928,0.16506805){\makebox(0,0)[lt]{\lineheight{1.25}\smash{\begin{tabular}[t]{l}$q_5$\end{tabular}}}}%
    \put(0.87912839,0.10672471){\makebox(0,0)[lt]{\lineheight{1.25}\smash{\begin{tabular}[t]{l}$m_1$\end{tabular}}}}%
    \put(0.92416376,0.11790378){\makebox(0,0)[lt]{\lineheight{1.25}\smash{\begin{tabular}[t]{l}$q_1$\end{tabular}}}}%
    \put(0.97904527,0.15381385){\makebox(0,0)[lt]{\lineheight{1.25}\smash{\begin{tabular}[t]{l}$g$\end{tabular}}}}%
    \put(0.85293767,0.17619419){\makebox(0,0)[lt]{\lineheight{1.25}\smash{\begin{tabular}[t]{l}$q_2$\end{tabular}}}}%
    \put(0.93406514,0.07797879){\makebox(0,0)[lt]{\lineheight{1.25}\smash{\begin{tabular}[t]{l}$u_1$\end{tabular}}}}%
    \put(0.91330416,0.15734961){\makebox(0,0)[lt]{\lineheight{1.25}\smash{\begin{tabular}[t]{l}$u_2$\end{tabular}}}}%
    \put(0.84926452,0.14361545){\makebox(0,0)[lt]{\lineheight{1.25}\smash{\begin{tabular}[t]{l}$m_2$\end{tabular}}}}%
    \put(0,0){\includegraphics[width=\unitlength,page=15]{dynsys.pdf}}%
  \end{picture}%
\endgroup%

%% file: chapters/chapter4.tex
\subsection{DISCUSSION}
\label{sec_discussion}
The results show that the DD-PINN successfully addresses the three limitations of the PINC as formulated in Sec.~\ref{sec_intro}: \ref{lim1} Without the need of automatic differentiation to calculate the physics loss, \emph{training times were reduced significantly for all three tested systems by a factor of 5--38} while evaluating 10--25 times more collocation points per epoch. A higher count of collocation points improves the coverage over the sampling ranges.
In the majority of cases, the DD-PINN also reached lower physics validation errors with a significant improvement for the two-link manipulator system that enabled \emph{stable self-loop prediction in contrast to the PINC}.
\ref{lim2} The limitations of the zero-order hold assumption are overcome by introducing the option for higher-order excitation. This was shown to be necessary for the five-mass-chain system to prevent induced vibrations from the interval-wise step excitation for $\delta_u=0$. \emph{Also, a higher-order excitation in combination with a longer shooting interval enables faster prediction while being more accurate} as shown for the nonlinear mass-spring damper system.
Lastly \ref{lim3}, the initial-condition loss for the DD-PINN is negligible ($\mathcal{L}_\mathrm{IC}\equiv 0$). While a loss-balancing scheme such as LRA may still be employed when the data loss is used, the \emph{balancing process is strongly simplified}. This is because the reduction of the physics loss towards a solution of the initial-value problem for the PINC depends on a sufficiently low initial-condition loss.

The DD-PINN therefore provides a solution to these limitations while being easy to implement and interchangeable to the classical PINC architecture. 
Also, it is compatible with a wide range of arbitrary Ansatz functions that may be chosen suitable for the problem at hand.

%% file: chapters/chapter5.tex
\section{CONCLUSIONS}
\label{sec_conclusion}
Physics-informed neural networks (PINNs) represent a well-established machine-learning framework that combines learning the solution of differential equations with supervised learning based on data. In its adaptation to dynamical systems, namely the PINN for control (PINC), we identified three primary limitations: \ref{lim1} Long training times for large and/or complex state-space models, \ref{lim2} limitations arising from the zero-order-hold assumption for excitation, and \ref{lim3} the need for hyperparameter-sensitive loss balancing schemes.

In this study, we introduced the domain-decoupled physics-informed neural network (DD-PINN) as a solution to these limitations.
We first formulated the DD-PINN architecture, showcasing how it enables calculation of gradients for the physics loss in closed form, is compatible to higher-order excitation input, and always has an initial-condition loss of zero. We then compared the DD-PINN to the PINC in simulation for three benchmark systems.
The results demonstrated that the DD-PINN significantly reduces training times while maintaining or surpassing the prediction accuracy of the PINC. Thereby, the self-loop prediction time of the DD-PINN is comparable to the PINC.

The DD-PINN allows for fast and accurate learning of large and complex dynamical systems, which were previously out of reach for the PINC. Its fast prediction abilities create opportunities for enabling MPC in larger dynamical systems, where traditional methods like numerical integrators are too slow, and training a PINC is not practical. Here, the data efficiency of physics-informed machine learning remains, making it possible to integrate sparse datasets with the system-governing physical equations.
Future work could explore using DD-PINN for higher-dimensional nonlinear systems to realize accurate state estimation or model predictive control.

%% file: tables/network_parameters.tex
\begin{table}[h]
\centering
\caption{PINC and DD-PINN training and neural-network parameters for the three evaluated systems.}
\label{tab_net_param}
\small
\begin{tabular}{lrrrrrr} \hline
                                   & \multicolumn{2}{c}{\textbf{Nonlinear mass-spring-damper}} & \multicolumn{2}{c}{\textbf{Five-mass chain}} & \multicolumn{2}{c}{\textbf{Two-link manipulator}} \\
                                    & PINC                    & DD-PINN                & PINC             & DD-PINN          & PINC               & DD-PINN             \\
\hline \hline
Epochs                              & $\leq\num{4000}$      & $\leq\num{4000}$     & $\num{2400}$     & $\num{2500}$     & $\num{7000}$       & $\num{2500}$        \\
Batches                             & $\num{400}$             & $\num{1500}$           & $\num{400}$      & $\num{500}$      & $\num{400}$        & $\num{1000}$        \\
$n_\mathrm{IC}$                     & $\num{40000}$           & 0                      & $\num{40000}$    & 0                & $\num{40000}$      & 0                   \\
$n_\mathrm{collo}$                  & $\num{20000}$           & $\num{375000}$          & $\num{20000}$    & $\num{250000}$   & $\num{20000}$      & $1 \cdot 10^6$      \\
$n_\mathrm{data}$ & $\num{2000}$            & $\num{2000}$           & 0                & 0                & 0                  & 0                   \\
Neurons                             & 96                      & 96                     & 128              & 128              & 64                & 128                 \\
Hidden layers                       & 1                       & 1                      & 2                & 2                & 4                  & 2 \\ \hline               
\end{tabular}
\end{table}

%% file: tables/results_tab1a.tex
\begin{table}[h]
\centering
\caption{PINC and DD-PINN training results for the nonlinear mass-spring-damper. Best values are highlighted in bold.}
\label{tab_results_msd}
\small
\begin{tabular}{l@{}rrr@{\hskip 0.5in}rrr}
\hline
& \multicolumn{6}{c}{\textbf{Nonlinear mass-spring-damper}} \\
 & \multicolumn{3}{c}{PINC} & \multicolumn{3}{c}{DD-PINN\textsuperscript{*}} \\
\begin{tabular}[c]{@{}l@{}}Number of\\ models trained\end{tabular} & 1 & 1 & 1 & 9 & 9 & 9 \\
$f$ in Hz & 50 & 100 & 200 & 50 & 100 & 200 \\
\hline
\hline
\begin{tabular}[r]{@{}l@{}}Training\\ time in h\end{tabular} & 8.04 & 9.22 & 9.70 & \begin{tabular}[r]{@{}l@{}}0.64\\ 1.65\end{tabular} & \begin{tabular}[r]{@{}l@{}}0.76\\ 1.70\end{tabular} & \begin{tabular}[r]{@{}l@{}}$\mathbf{0.54}$\\ 1.61\end{tabular} \\
$\mathcal{L}_\mathrm{IC}$ & $5.20 \cdot 10^{-7}$ & $1.67 \cdot 10^{-7}$ & $1.11 \cdot 10^{-7}$ & \begin{tabular}[r]{l}$\mathbf{0}$\\$\mathbf{0}$\end{tabular} & \begin{tabular}[r]{l}$\mathbf{0}$\\$\mathbf{0}$\end{tabular} & \begin{tabular}[r]{l}$\mathbf{0}$\\$\mathbf{0}$\end{tabular} \\
$\mathcal{L}_\mathrm{phys}$ & $2.26 \cdot 10^{-5}$ & $2.05 \cdot 10^{-6}$ & $3.72 \cdot 10^{-7}$ & \begin{tabular}[r]{@{}l@{}}$2.90 \cdot 10^{-5}$\\ $7.67 \cdot 10^{-5}$\end{tabular} & \begin{tabular}[r]{@{}l@{}}$1.08 \cdot 10^{-6}$\\ $4.87 \cdot 10^{-6}$\end{tabular} & \begin{tabular}[r]{@{}l@{}}$\mathbf{5.49 \cdot 10^{-8}}$\\ $2.42 \cdot 10^{-7}$\end{tabular} \\
$\mathcal{L}_\mathrm{data}$ & $1.38 \cdot 10^{-4}$ & $1.54 \cdot 10^{-5}$ & $4.55 \cdot 10^{-7}$ & \begin{tabular}[r]{@{}l@{}}$1.21 \cdot 10^{-8}$\\ $1.78 \cdot 10^{-4}$\end{tabular} & \begin{tabular}[r]{@{}l@{}}$6.79 \cdot 10^{-10}$\\ $1.99 \cdot 10^{-5}$\end{tabular} & \begin{tabular}[r]{@{}l@{}}$\mathbf{1.15 \cdot 10^{-11}}$\\ $7.83 \cdot 10^{-8}$\end{tabular} \\
\begin{tabular}[r]{@{}l@{}}Scaled MSE\\ on test path\end{tabular} & $1.93 \cdot 10^{-1}$ & $1.92 \cdot 10^{-2}$ & $1.92 \cdot 10^{-3}$ & \begin{tabular}[r]{@{}l@{}}$1.53 \cdot 10^{-3}$\\ $2.40 \cdot 10^{-1}$\end{tabular} & \begin{tabular}[r]{@{}l@{}}$4.30 \cdot 10^{-4}$\\ $3.14 \cdot 10^{-2}$\end{tabular} & \begin{tabular}[r]{@{}l@{}}$\mathbf{2.26 \cdot 10^{-4}}$\\ $4.07 \cdot 10^{-3}$\end{tabular} \\
\begin{tabular}[r]{@{}l@{}}Relative pred.\\ time\end{tabular} & $\mathbf{0.43}$ & 1.00 & 1.62 & \begin{tabular}[r]{@{}l@{}}0.57\\ 0.67\end{tabular} & \begin{tabular}[r]{@{}l@{}}1.15\\ 1.26\end{tabular} & \begin{tabular}[r]{@{}l@{}}2.34\\ 2.79\end{tabular} \\
\hline
\end{tabular}
\\
\textsuperscript{*}Values denote min and max values among the different configurations with varying degree of excitation $\delta_u \in \{0, 1, 2\}$\\ and different numbers of base construction functions $n_g \in \{5, 20, 50\}$.
\end{table}

%% file: tables/results_tab1b.tex
\begin{table}[h]
\centering

\caption{PINC and DD-PINN training results for the five-mass-chain and two-link manipulator systems. Best values per system are highlighted in bold.}    
\label{tab_results_other}
\small
\begin{tabular}{lrr@{\hskip 0.5in}rr}
\hline
 & \multicolumn{2}{c}{\textbf{Five-mass chain}} & \multicolumn{2}{c}{\textbf{Two-link manipulator}} \\
 & PINC & DD-PINN\textsuperscript{*} & PINC & DD-PINN \\
\hline
\hline
\begin{tabular}[r]{@{}l@{}}Training\\ time in h\end{tabular} & \begin{tabular}[r]{l}144.5\end{tabular} & \begin{tabular}[r]{l}4.29\\ $\mathbf{3.75}$\end{tabular} & \begin{tabular}[r]{l}211.0\end{tabular} & \begin{tabular}[r]{l}$\mathbf{7.41}$\end{tabular} \\
$\mathcal{L}_\mathrm{IC}$ & \begin{tabular}[r]{l}$8.80 \cdot 10^{-5}$\end{tabular} & \begin{tabular}[r]{l}$\mathbf{0}$\\$\mathbf{0}$\end{tabular} & \begin{tabular}[r]{l}$8.90 \cdot 10^{-6}$\end{tabular} & \begin{tabular}[r]{l}$\mathbf{0}$ \\ $\mathbf{0}$\end{tabular} \\
$\mathcal{L}_\mathrm{phys}$ & \begin{tabular}[r]{l}$\mathbf{6.27 \cdot 10^{-3}}$\end{tabular} & \begin{tabular}[r]{l}$8.43 \cdot 10^{-3}$ \\ $8.22 \cdot 10^{-3}$\end{tabular} & \begin{tabular}[r]{l}$2.46 \cdot 10^{-2}$\end{tabular} & \begin{tabular}[r]{l}$\mathbf{3.22 \cdot 10^{-3}}$\end{tabular} \\
$\mathcal{L}_\mathrm{data}$ &  &  &  &  \\
\begin{tabular}[r]{@{}l@{}}Scaled MSE\\ on test path\end{tabular} & \begin{tabular}[r]{l}$1.78 \cdot 10^{-1}$\end{tabular} & \begin{tabular}[r]{l}$2.42 \cdot 10^{-1}$\\ $\mathbf{1.27 \cdot 10^{-3}}$\end{tabular} & \begin{tabular}[r]{l}$9.17 \cdot 10^{-2}$\end{tabular} & \begin{tabular}[r]{l}$\mathbf{1.68 \cdot 10^{-3}}$\end{tabular} \\
\begin{tabular}[r]{@{}l@{}}Relative pred. time\end{tabular} & \begin{tabular}[r]{l}$\mathbf{1.00}$\end{tabular} & \begin{tabular}[r]{l}1.33\\ 1.34\end{tabular} & \begin{tabular}[r]{l}$\mathbf{1.00}$\end{tabular} & \begin{tabular}[r]{l}1.03\end{tabular} \\
\hline
\end{tabular}
\\
\textsuperscript{*}Values for the DD-PINN for the five-mass-chain system correspond to the configurations with $\delta_u=0$ and $\delta_u=2$.
\end{table}